\documentclass[11pt]{article}
\usepackage{amsmath,amssymb,color,graphics,epsfig}

\usepackage{amsfonts}

\newcommand{\be}{\begin{equation}}
\newcommand{\ee}{\end{equation}}
\newcommand{\bea}{\setlength\arraycolsep{2pt} \begin{eqnarray}}
\newcommand{\eea}{\end{eqnarray}}
\newcommand{\nn}{\nonumber}

\def\ft#1#2{{\textstyle{\frac{\scriptstyle #1}{\scriptstyle #2} } }}
\def\fft#1#2{{\frac{#1}{#2}}}

\def\0{{\sst{(0)}}}
\def\1{{\sst{(1)}}}
\def\2{{\sst{(2)}}}
\def\3{{\sst{(3)}}}
\def\4{{\sst{(4)}}}
\def\5{{\sst{(5)}}}
\def\6{{\sst{(6)}}}
\def\7{{\sst{(7)}}}
\def\8{{\sst{(8)}}}
\def\sst#1{{\scriptscriptstyle #1}}

\begin{document}

\begin{flushright}
\end{flushright}

\vspace{25pt}
\begin{center}
{\large {\bf Static and Dynamic Hairy Planar Black Holes}}

\vspace{10pt}
Zhong-Ying Fan and H. L\"u

\vspace{10pt}

{\it Center for Advanced Quantum Studies\\ }
{\it Department of Physics, Beijing Normal University, Beijing 100875, China}

\vspace{40pt}

\underline{ABSTRACT}
\end{center}

We consider Einstein gravity in general dimensions, coupled to a scalar field either minimally or non-minimally, together with a generic scalar potential.  By making appropriate choices of the scalar potential, we obtain large classes of new scalar hairy black holes that are asymptotic to anti-de Sitter spacetimes in planar coordinates.  For some classes of solutions, we can promote the scalar charge to be dependent on the advanced or retarded times in the Eddington-Finkelstein coordinates, and obtain exact dynamic solutions.  In particular, one class of the collapse solutions describe the evolution from the AdS vacua to some stable black hole states, driven by a conformally-massless scalar.  It is an explicit demonstration of nonlinear instability of the AdS vacuum that is stable at the linear level.

\vfill {\footnotesize Emails: zhyingfan@gmail.com \ \ \ mrhonglu@gmail.com}

\thispagestyle{empty}

\pagebreak

\tableofcontents
\addtocontents{toc}{\protect\setcounter{tocdepth}{2}}




\section{Introduction}

Owing to the AdS/CFT correspondence \cite{mald,gkp,wit}, the construction of classical solutions in general relativity can have applications in quantum field theories.  One particularly interesting class of solutions is about the dynamic formation of black holes that are asymptotic to anti-de Sitter (AdS) spacetimes.  They can be used to study non-equilibrium thermalization of certain strongly-coupled field theory \cite{Bhattacharyya:2009uu}.  Black hole formation is an old subject and it can be addressed within the Robinson-Trautman ansatz \cite{Robinson:1960zzb,Robinson:1962zz}.  However, few examples of exact solutions have been found for a given Lagrangian. Restricting to spherical symmetries, an exact solution of a charged radiating white hole was obtained in $N=4$, $D=4$ supergravity \cite{Gueven:1996zm}.  The white hole eventually settles down smoothly to a black hole state at late retarded times.

Recently a first example of exact solution of black hole formation was constructed in some Einstein-scalar gravity in four dimensions \cite{Zhang:2014sta}.  Analogous to the example in \cite{Gueven:1996zm}, the formation is driven by the non-conserved scalar charge.  It eventually settles down to a black hole state at late advanced times with the relaxation time that is inversely proportional to the mass of the final black hole.  The solution was later generalized to include charges \cite{Lu:2014eta} and the general solutions contain the radiating white hole of \cite{Gueven:1996zm} as a special example.  Further properties and global structures were discussed in \cite{Zhang:2014dfa}. Analogous black hole formation in three dimensions was also found in \cite{Xu:2014xqa}.  Exact solutions of dynamic C-metrics in STU gauged supergravity model and beyond were constructed in \cite{Lu:2014ida,Lu:2014sza}.

One important motivation for us to construct dynamic solution is to study the stability of the AdS vacua.  It is well known that the AdS spacetime with a massive scalar that satisfies the Breitenlohner-Freedman bound is stable at the linear level.  However, numerical studies indicate that nonlinear instability exists and the vacuum can evolve into a final black hole state \cite{Bizon:2011gg,Buchel:2012uh,Wu:2013qi,Buchel:2013uba}.  We would like to construct solutions that demonstrate such nonlinear instability analytically.

  The purpose of this paper is to construct more examples of such exact formations of black holes in general dimensions.  However, we restrict our attention to Einstein-scalar gravities.  Furthermore, for future applications in the AdS/CFT correspondence, we would like to construct solutions that are asymptotic to the AdS in planar coordinates, so that the boundary is the Minkowski spacetime.  The corresponding black objects in the bulk are called AdS planar black holes.  Our strategy is follows. We start with Einstein gravity coupled to a scalar with minimal or non-minimal couplings.  We choose the scalar potential appropriately so that we obtain a large number of exact solutions of static AdS planar black holes carrying a scalar charge.  We then follow \cite{Zhang:2014sta} and rewrite the static solutions in Eddington-Finkelstein coordinates and promote the scalar charge to be dependent on the advanced or retarded times. This procedure of \cite{Zhang:2014sta} does not guarantee to succeed. In fact, for the majority of the static solutions we shall construct, the procedure yields no further solutions.  However, we do obtain two classes of new exact dynamic solutions in all $n\ge 3$ dimensions.

The paper is organized as follows. In section 2, we discuss the Einstein gravity coupled non-minimally to a scalar field, together with a generic scalar potential.  In section 3, we construct large classes of scalar hairy AdS planar black holes in the non-minimally coupled theories.  We analyse the global structure and obtain the corresponding first law of thermodynamics.  In section 4, we obtain a class of exact dynamic solutions in general dimensions.  We show that the solutions describe black hole formation due to the nonlinear instability of the AdS vacua.  In section 5, we consider Einstein gravity with a minimally-coupled scalar in general dimensions and obtain two new classes of scalar hairy AdS planar black holes.  We then promote these solutions to be time-dependent and obtain a new class of radiating white holes that eventually decay to become AdS vacua.  We conclude the paper in section 6.

\section{Einstein gravity with a non-minimally coupled scalar}

We begin with Einstein gravity in general $n\ge 3$ dimensions, coupled non-minimally to a scalar with a generic scalar potential. The Lagrangian is:
\be
\mathcal{L}_n=\sqrt{-g}\Big(\kappa_0 R-\ft12\xi\phi^2 R
-\ft12 (\partial \phi)^2-V(\phi) \Big) \label{gen-nonmin-lag}\,,
\ee
where $\kappa_0$ is related to the inverse of the bared Newton's constant, and
$\xi$ is a constant that characterizes the coupling strength between the scalar $\phi$ and the curvature. In $n$ dimensions, the scalar sector with vanishing potential $V(\phi)$ becomes conformally invariant when $\xi$ takes the value of
\be
\xi=\frac{n-2}{4(n-1)}\,.
\ee
We shall let $\xi$ be general in this paper unless specified otherwise. Since we consider non-vanishing $\kappa_0$ throughout the paper, we may let, without loss of generality,
\be
\kappa_0=\ft12\xi\,.
\ee
The effective Newton's ``constant'' now becomes scalar dependent, given by the inverse of $\kappa(\phi)$, defined by
\be
\kappa(\phi)=\kappa_0 - \ft12\xi \phi^2=\ft12\xi (1-\phi^2)\,.
\ee
We require that $\kappa(\phi)$ be positive on and outside the event horizon of a static black hole in order to avoid ghost-like graviton modes.

The covariant equations of motion are
\be
E_{\mu\nu}\equiv\kappa_0 G_{\mu\nu}-T_{\mu\nu}^{\rm (min)}-T_{\mu\nu}^{\rm (non)}\,,\qquad
\Box\phi = \xi\phi R + \fft{dV}{d\phi}\,,\label{non-min-geneom}
\ee
where $ G_{\mu\nu}=R_{\mu\nu}-\fft 12 R g_{\mu\nu}$ is the Einstein tensor and
\bea
T_{\mu\nu}^{\rm (min)}&=&\ft 12\partial_\mu\phi \partial_\nu \phi-\ft 12 g_{\mu\nu}\Big(\ft 12(\partial \phi)^2+V(\phi) \Big)\,,\nn\\
T_{\mu\nu}^{\rm (non)}&=&\ft12\xi(\phi^2 G_{\mu\nu}+g_{\mu\nu}\Box \phi^2-\nabla_\mu \nabla_\nu \phi^2)\,.
\eea
In this paper, we require that the scalar potential have a stationary point at $\phi=0$, with a negative cosmological constant $V(0)<0$.  We would also like that the scalar's effective mass in the corresponding AdS vacuum satisfies the Breitenlohner-Freedman bound, so that the vacuum is stable at the linear level.

\section{Static AdS planar black holes}

The construction of black holes from the theories (\ref{non-min-geneom}) has been undertaken previously \cite{Martinez:1996gn,Nadalini:2007qi,Anabalon:2012ta,Zou:2014gla}. In this section, we construct a much more general class of new exact planar black holes to (\ref{non-min-geneom}).  We consider a specialized ansatz
\be
ds_n^2 = -f(r) dt^2 + \fft{dr^2}{f(r)} + r^2\,dx^i dx^i\,,\qquad
\phi=\phi(r)\,.\label{ansatz1}
\ee
In the most general ansatz that respects the isometry, $g_{tt}$ and $g_{rr}$ are instead two {\it independent} functions of $r$.  In the usual construction of such solutions, one tries to solve for these two functions and $\phi$ for a given potential $V$. It is very unlikely to find an exact solution for a random potential $V$.  The purpose of this paper is not trying to find the most general solution for a given scalar potential $V(\phi)$, but instead, we make a simplified metric ansatz (\ref{ansatz1}) so that $f$ and $\phi$ can be solved, independent of $V$. The scalar potential is then treated as a unknown function that can be finely tuned and explicitly derived.

   There are three independent Einstein equations, and we have three unknown functions, $f$, $\phi$ and $V$.  In particular, $V$ appears in Einstein equations linearly and algebraically. We can thus eliminate $V$ and obtain two differential equations for $f$ and $\phi$.  In particular, eliminating $V$ in the $E_{tt}=0$ and $E_{rr}=0$ equations yields
\be
\xi \phi \phi'' + (\xi-\ft12) \phi'^2=0\,.
\ee
Note that this equation is independent of the spacetime dimension and also the topology of level surfaces.  It can be solved exactly, given by
\be
\phi=\fft{1}{(c_1 r + c_2)^\mu}\,,\qquad \xi=\fft{\mu}{2(2\mu+1)}\,,\label{scalarsol}
\ee
where $c_1$ and $c_2$ are two integration constants, and $\mu$ is a reparametrization of $\xi$.  We require that $\mu>0$ so that the scalar vanishes at asymptotic infinity.  The remaining Einstein equation is then a second-order linear differential equation for $f$.  For appropriate dimension $n$ and $\mu$, it can be solved explicitly. One can then obtain the scalar potential $V$ as a function of $r$ from the remainders of the Einstein equations.  Owing to the relation (\ref{scalarsol}), $V$ can be expressed as a function of scalar $\phi$.  Interestingly, the scalar equation is automatically satisfied. Thus one obtains an exact solution associated with the scalar potential.  However, there is a possible short coming for this procedure.  All the integration constants, $(c_1,c_2)$ and two more from the $f$ equation, may in general appear in the derived scalar potential as well.  It follows that the solution has no free parameter since the constants in the solution are all coupling constants of the theory.

    For our purpose, we are interested in a solution with at least one free parameter, which we allow to vary with time in the dynamic generalization.  It turns out that our ansatz (\ref{ansatz1}) indeed has this property and each black hole solution has the scalar charge as a free parameter.  We would like present the results in three cases, namely (1) $c_2=0$, (2) $c_2=1$ and (3) $c_2$ is generic.

\subsection{Case 1: $c_2=0$}

This is the simplest case. We may reparameterize the scalar solution as
\be
\phi=\Big(\fft{q}{r}\Big)^\mu\,,\label{phires1}
\ee
where the constant $q$ can be viewed as the scalar charge.  We find that the second-order linear differential equation for $f$ is given by
\be
r^2(r^{2\mu}-q^{2\mu}) f'' +r\big( (n-4) r^{2\mu} -(n-2\mu-4)q^{2\mu} \big) f' -
2\big((n-3) r^{2\mu} -(n-2\mu-3)q^{2\mu}\big) f=0\,.\label{feom1}
\ee
It can be solved exactly in terms of a hypergeometric function
\be
f=g^2 r^2 -\alpha \ft{ q^{n-1}}{r^{n-3}}\, {}_2F_1[1,\ft{n-1}{2\mu};1+\ft{n-1}{2\mu};
\ft{q^{2\mu}}{r^{2\mu}}]\,.\label{fres1}
\ee
Here $g$ and $\alpha$ are two integration constants.  When $\alpha=0$, the metric becomes the AdS spacetime in planar coordinates with radius $\ell=1/g$:
\be
ds^2 = \fft{dr^2}{g^2 r^2} + r^2 (-g^2 dt^2 + dx^i dx^i)\,,\label{adsvac}
\ee
even though the scalar is non-vanishing.  This is characteristic in a theory with a non-minimally coupled scalar.  For non-vanishing $\alpha$, which we consider in this paper, the AdS vacuum emerges when we turn off the scalar charge, namely $q=0$.

The scalar potential can be determined by the following equation
\bea
V&=&-\fft{\mu}{4(2\mu+1)}\Big(\big(n-2 -(n-2\mu-2)(\ft{q}{r})^{2\mu}\big)
\fft{f'}{r}\cr
&&+\big( (n-2)(n-3)-(n-2\mu-2)(n-2\mu-3)(\ft{q}{r})^{2\mu}\big) \fft{f}{r^2}\Big)\,.
\eea
Substituting (\ref{fres1}) and also making use of (\ref{phires1}), we obtain the scalar potential
\bea
V &=&\ft{\mu g^2}{4(2\mu+1)}\Big((n-2\mu-1)(n-2\mu-2)\phi^2 - (n-1)(n-2)\Big)\cr
&&+\ft{\alpha\mu \phi^\fft{n-1}{\mu}}{4(2\mu+1)(1-\phi^2)}\Big(
(n-1)((n-2\mu-2)\phi^2-n+2)\cr
&&\qquad+(1-\phi^2)\big((n-1)(n-2) - (n-2\mu-1)(n-2\mu-2)\phi^2\big)\cr
&&\qquad\qquad\times {}_2F_1[1,\ft{n-1}{2\mu};1+\ft{n-1}{2\mu};
\phi^2]\Big)\,.
\eea
Note that the scalar potential has three parameters $(g,\alpha,\mu)$, but it is independent of $q$.
The scalar potential has a stationary point at $\phi=0$, with
\be
V(0)=-(n-1)(n-2)\kappa_0 g^2=-\fft{(n-1)(n-2) \mu g^2}{4(2\mu+1)}\,.
\ee
Thus the metric (\ref{adsvac}) is the AdS vacuum of the theory. The linearized scalar equation around the AdS vacuum implies that the mass of the scalar is given by
\be
m^2=-(n-1-\mu)\mu g^2=-\ft14(n-1)^2 g^2 + \ft14 (n-1-2\mu)^2 g^2\,.
\ee
Thus the Breitenlohner-Freedman bound $m^2\ge- \ft14(n-1)^2 g^2$ is always satisfied.  Consequently, the AdS vacuum is linearly stable against perturbations.

We now show that the solutions (\ref{ansatz1}) with (\ref{fres1}) describe AdS (planar) black holes in general dimensions.  First we note that the asymptotic behavior of the function $f$ is
\be
f=g^2 r^2 - \alpha q^2 \Big(\fft{q}{r}\Big)^{n-3} \Big(1 + \ft{n-1}{n+2\mu-1}
\Big(\fft{q}{r}\Big)^{2\mu} + \ft{n-1}{n+4\mu-1}
\Big(\fft{q}{r}\Big)^{4\mu} + \cdots\Big)\,.
\ee
We can thus read off the mass of the solution
\be
M=\fft{(n-2)\kappa_0 \alpha}{16\pi} q^{n-1}\,.\label{mass1}
\ee
The positiveness of the mass requires that $\alpha>0$.  This is a general conclusion in this paper and we shall not emphasize this further. In order to establish that there is a horizon, we note that
\be
\lim_{r\rightarrow q+} {}_2F_1[1,\ft{n-1}{2\mu};1+\ft{n-1}{2\mu};
\ft{q^{2\mu}}{r^{2\mu}}]=+\infty\,.
\ee
It follows that there must exist a certain $r=r_0>q$ for which $f(r_0)=0$, corresponding to the event horizon.  The horizon exists regardless how small $q$ is as long as it is non-vanishing and positive. This is analogous to the Schwarzschild black hole.  Interestingly for our solution, we have
\be
\kappa(\phi) = \fft{\mu}{4(2\mu+1)} \Big(1 - \Big(\fft{q}{r}\Big)^{2\mu}\Big)\,,
\ee
which is positive definite for the region on and outside of the event horizon $r\ge r_0>q$. The temperature and the entropy density are given by
\be
T=\fft{(n-1)\alpha q^{n-1}}{4\pi r_0^{n-2\mu-2}(r_0^{2\mu} - q^{2\mu})}\,,\qquad
S=\ft14\kappa_0 r_0^{n-2\mu-2} (r_0^{2\mu}-q^{2\mu})\,.
\ee
Note that $f(r_0)=0$ implies that $r_0/q$ is a pure numerical number, and hence we have
\be
dr_0=\fft{r_0}{q} dq\,.\label{r0qrelation}
\ee
It follows that the first law of thermodynamics
\be
dM=TdS\label{firstlaw}
\ee
holds straightforwardly.  The mass, temperature and entropy satisfy the Smarr relation
\be
M=\fft{n-2}{n-1} T S\,.\label{smarr}
\ee

\subsection{Case 2: $c_2=1$}

We reparameterize the $\phi$ solution as
\be
\phi=\fft{q^\mu}{(r+q)^\mu}\,.
\ee
For this choice, the effective gravitational coupling is given by
\be
\kappa(\phi) = \fft{\mu}{4(2\mu+1)}\Big(1 - \big(\ft{q}{r+q}\big)^{2\mu}\Big)\,,
\ee
which is positive definite for positive $r$.  The scalar potential for this system can be determined by
\bea
V &=& \fft{\mu}{4(2\mu+1)}\Big[\Big(2-n+\ft{(n+2\mu-2)r +(n-2)q}{r+q} \big(\ft{q}{r+q}\big)^{2\mu}\Big)\fft{f'}{r}\cr
&&+\Big((n-2)(n-3)\big(1 - \big(\ft{q}{r+q}\big)^{2(1+\mu)}\big) \cr
&&\qquad-\ft{(n-2\mu-3)r((n-2\mu-2) r + 2(n-2)q)}{(r+q)^2}\big(\ft{q}{r+q}\big)^{2\mu}\Big)\fft{f}{r^2}\Big]\,.
\eea
The function $f$ satisfies a second-order linear differential equation:
\bea
0 &=& r^2(r+q)\big((r+q)^{2\mu}-q^{2\mu}\big) f'' \cr
&&+r\big( (n-4) (r+q)^{1+2\mu} -
((n-2\mu-4)r +(n-4)q) q^{2\mu}\big) f'\cr
&&-2\big((n-3)(r+q)^{1+2\mu} - ( (n-2\mu-3) r + (n-3) q) q^{2\mu}\big) f\,.
\eea
For $\mu$ being half integers, the coefficients of the above linear differential equation are polynomials of $r$.  The equation can be solved for low-lying half integer $\mu$.

For example, when $\mu=\ft12$, we have a simple solution in general dimensions $n\ge 3$, given by
\bea
V &=& -\ft1{16}g^2 (1-\phi^2) \big(2\phi^4 + 2 (n-2)\phi^2 + (n-1)(n-2)\big) -
\fft{\alpha \phi^{2n}}{8n (1-\phi^2)^{n-3}}\,,\cr
f &=& g^2 r^2 - \alpha q^2 \Big(\big(\fft{q}{r}\big)^{n-3} + \ft{n-1}{n}\big(\fft{q}{r}\big)^{n-2}\Big)\,.
\eea
The solution is asymptotic to AdS in planar coordinates with a horizon at some $r_0>0$ with $f(r_0)=0$. The mass density is given by (\ref{mass1}).  The temperature and the entropy density are
\be
T=\fft{(n-1)\alpha (r_0+q)}{4\pi} \big(\fft{q}{r_0}\big)^{n-1}\,,\qquad
S=\fft{\kappa_0 r_0^{n-1}}{4(r_0+q)}\,.
\ee
It follows from (\ref{r0qrelation}) that the first law of thermodynamics (\ref{firstlaw}) and the Smarr relation (\ref{smarr}) are both satisfied.  This $\mu=\fft12$ solution in $n=3$ dimensions was previously obtained in \cite{Nadalini:2007qi}.

For $\mu=1$, the results are somewhat more complicated, and we have
\bea
V&=& -\ft1{12} g^2 (1-\phi)\big(6\phi^3 + 2(2n-5)\phi^2 + (n-1)(n-2)\phi +
(n-1)(n-2)\big)\cr
&&-\fft{\alpha\,\phi^n}{24n(1+\phi)(1-\phi)^{n-1}}\Big[6(n+1)\phi^4 +4(n^2-3n-1)\phi^3\cr
&&+(n^3-6n^2+n-8)\phi^2 -2(n^2-5n-2)\phi - (n-1)^2(n-2)\cr
&&+(n-1)(1-\phi)\big(6\phi^3 + 2(2n-5)\phi^2 + (n-1)(n-2)(1+\phi)\big)
{}_2F_1[1,1;n+1;\ft{2\phi}{1+\phi}]\Big]\,,
\cr
f&=&g^2 r^2 -\fft{\alpha q^{n-1}}{r^{n-3}}\Big(1 + \fft{(n-1)q}{2nr} -
\fft{(n-1)q}{2n(r+2q)}{}_2F_1[1,1;n+1;\ft{2q}{r+2q}]\Big)\,.
\eea
The mass is again (\ref{mass1}) and the temperature and the entropy density are given by
\be
T=\fft{(n-1)\alpha(r_0+q)^2}{4\pi(r_0+2q)} \big(\fft{q}{r_0}\big)^{n-1}\,,\qquad
S=\fft{\kappa_0 r_0^{n-1}(r_0+2q)}{4(r_0+q)^2}\,.
\ee
The first law of thermodynamics (\ref{firstlaw}) and the Smarr relation (\ref{smarr}) can be easily shown to hold.

Exact solutions can also be obtained for $\mu=\ft32$ and $2$ for low dimensions and we shall not enumerate them here.

\subsection{Case 3: general $c_2\ne 0$}

Solutions for general $(c_1,c_2)$ parameters are also possible.  We first reparameterize the scalar solution (\ref{scalarsol}) as
\be
\phi=\fft{q^\mu}{(r+ \gamma q)^\mu}\,.
\ee
Here we have replaced the constants $(c_1,c_2)$ by the scalar charge $q$ and a new parameter $\gamma$. Following the same steps of the earlier examples, we obtain for $\mu=\ft12$
\be
f=g^2r^2 - \alpha q^2\Big(\big(\fft{q}{r}\big)^{n-3} + \ft{n-1}{n}\big(\fft{q}{r}\big)^{n-2}\,
{}_2F_1[1,n;n+1;\fft{(1-\gamma)q}{r}]\Big)\,.
\ee
The associated scalar potential is given by
\bea
V &=&\ft{1}{16}\Big(2\gamma^2 \phi^6 + 2(n-3)\gamma \phi^4 + (n-2)(n-3)\phi^2 -
(n-1)(n-2)\Big)\Big[g^2 \cr
&&-\fft{\alpha (n-1)\phi^{2n}}{n(1-\gamma\phi^2)^n} {}_2F_1[1,n;1+n;
\fft{(1-\gamma)\phi^2}{1-\gamma\phi^2}]\Big]\cr
&&+ \fft{\alpha\phi^{2n}}{16(1-\phi^2)(1-\gamma\phi^2)^{n-1}}\Big(
2\gamma^2 \phi^6 + 2(n-\gamma-3)\gamma \phi^4\cr
&& + ((n-2)(n-3)-(n-5)\gamma)\phi^2
-n^2 + 4n-5\Big)\,.
\eea
Again, the scalar charge $q$ does not appear in the potential and hence it is a free parameter of the solutions. It can be calculated that these solutions describe AdS planar black holes with mass density (\ref{mass1}) and the temperature and entropy density
\be
T=\fft{(n-1)\alpha q (r_0 + \gamma q)}{4\pi (r_0 + (\gamma-1)q)} \big(\fft{q}{r_0}\big)^{n-2}\,,\qquad
S=\fft{\kappa_0(r_0+ (\gamma-1) q)}{4(r_0 + \gamma q)}r_0^{n-2}\,,
\ee
which satisfy the first law of thermodynamics (\ref{firstlaw}) and the Smarr relation (\ref{smarr}).

    For $\mu=1$, we find that the scalar potential is given by
\bea
V&=&\ft1{12} \Big(6\gamma^2 \phi^4 + 4(n-4)\gamma\phi^3 + (n-3)(n-4)\phi^2 -
(n-1)(n-2)\Big)\cr
&&\times\Big[g^2 +\fft{\alpha(n-1)\phi^n}{2n(1+\phi)(1-\gamma\phi)^{n-1}}\,
{}_2F_1[1,1;n+1; \fft{(1+\gamma)\phi}{1+\phi}]\cr
&&-\fft{\alpha(n-1)\phi^n}{2n(1-\phi)(1-\gamma\phi)^{n-1}}\,
{}_2F_1[1,1;n+1;-\fft{(1-\gamma)\phi}{1-\phi}]\Big]\cr
&& +\fft{\alpha\phi^{n+1}}{12(1-\phi^2)(1-\gamma\phi)^{n-1}}
\Big(6\gamma^2 \phi^4 + 4(n-4)\gamma\phi^3\cr
&& + (n^2-7n-6\gamma^2+12)\phi^2 - 2(n-7)\gamma \phi -n^2+5n-10\Big)\,.
\eea
The metric function $f$ is given by
\bea
f&=&g^2 r^2 -\alpha q^2 \big(\fft{q}{r}\big)^{n-3} -
\fft{\alpha (n-1) q^n}{2n(r+(\gamma-1)q)r^{n-3}}\, {}_2F_1[1,1,n+1,\fft{(\gamma-1)q}{r+(\gamma-1) q}]\cr
&& +
\fft{\alpha (n-1) q^n}{2n(r+(\gamma+1)q)r^{n-3}}\, {}_2F_1[1,1,n+1,\fft{(\gamma+1)q}{r+(\gamma+1) q}]\,.
\eea
The solutions describe AdS planar black holes with mass density (\ref{mass1}) and the temperature and entropy density
\be
T=\fft{\alpha (n-1) q^{n-1} r_0^{2-n} (r_0 + \gamma q)^2}{4\pi(r_0+(\gamma+1)q)
(r_0 + (\gamma-1)q)}\,,\qquad
S=\fft{\kappa_0(r_0 + (\gamma+1)q) (r_0 + (\gamma-1)q)}{4 (r_0+\gamma q)^2} r_0^{n-2}\,,
\ee
which satisfy the Smarr relation (\ref{smarr}) and the first law of thermodynamics (\ref{firstlaw}).

It is worth pointing out that the hypergeometric functions actually reduce to simpler expressions since $n$ is an integer.  For example, we have
\be
{}_2F_1[1,1;n+1,x]=-n(x-1)^{n-1} x^{-n} \log(1-x) + x^{1-n} P_{n-2} (x)\,.
\ee
where $P_m(x)$ denotes certain $x$ polynomial of order $m$. It is also clear that solutions of cases 1 and 2 are some subtle singular limits $\gamma=0$ and 1 of the most general case-3 solutions.

\section{Exact formation of AdS black holes}

Having obtained a large class of static AdS planar black holes in Einstein gravity with a non-minimally coupled scalar in general dimensions, we would like to construct their dynamic counterparts. A procedure of promoting the static black hole solution to be time-dependent was given in \cite{Zhang:2014sta}, and was discussed in the introduction. To be specific, we begin by rewriting the static ansatz (\ref{ansatz1}) in the Eddington-Finkelstein coordinates, namely
\be
ds^2 = 2 du dr - f(r,q) du^2 + r^2 dx^i dx^i\,,\qquad \phi=\phi(r,q)\,,
\ee
where $u=t + \int f^{-1} dr$ is the advanced time.  We then promote the scalar charge $q$, which is not conserved, to be a function of $u$, namely $a(u)$.  We then substitute the isotropic but time-dependent ansatz back into the equations of motion. Unfortunately, for the majority of the static solutions we have obtained in the previous section, the generalization does not solve the equations of motion. However, we find that there exists a class of static solutions in all $n\ge 3$ that do give rise to dynamic AdS black hole formations where $a$ satisfies a single second-order differential equation.

To be precise, for the case 1 solutions given in section 3.1, the above procedure indeed works for a specific choice of the solutions, namely
\be
\xi= \fft{n-2}{4(n-1)}\,.\label{conformalxi}
\ee
For this choice of $\xi$, the scalar sector with vanishing potential $V(\phi)$ becomes conformally invariant. Correspondingly the theory is given by
\bea
e^{-1}{\cal L}_n &=& \ft{n-2}{8(n-1)}(1-\phi^2)\, R - \ft12 (\partial\phi)^2 -
V\,,\cr
V&=&-\ft18(n-2)^2 \Big(g^2 + \alpha \phi^{\fft{2(n-1)}{n-2}}\big(\ft{1}{1-\phi^2}-
{}_2F_1[1,\ft{n-1}{n-2};\ft{2n-3}{n-2};\phi^2]\big)\Big)\,.
\eea
The scalar potential has a stationary point $\phi=0$.  The potential in small $\phi$ expansion is given by
\be
V=-\ft18(n-2)^2 g^2- \fft{(n-2)^3\alpha}{8(2n-3)} \phi^{4 + \fft{2}{n-2}} -
\fft{(n-2)^3\alpha}{4(3n-5)} \phi^{6 + \fft{2}{n-2}} + \cdots\,.
\ee
Thus the scalar potential gives no contribution to the mass of the scalar.  It follows from (\ref{conformalxi}) that the scalar $\phi$ is conformally massless, indicating that the AdS vacua is stable against linear perturbation.

The dynamic solutions are given by
\bea
ds^2 &=& 2du dr - f du^2 + r^2 dx^i dx^i\,,\qquad \phi=\big(\fft{a}{r}\big)^{\fft12(n-2)}\,,\cr
f&=& g^2 r^2 -\fft{\alpha a^{n-1}}{r^{n-3}}\,{}_2F_1[1,\ft{n-1}{n-2};
\ft{2n-3}{n-2}, \big(\ft{a}{r}\big)^{n-2}]\,,
\eea
where $a$ is function of $u$ only, satisfying the second-order nonlinear differential equation
\be
\fft{\ddot a}{a^2} - \fft{2\dot a^2}{a^3} + \fft{(n-1)\alpha \dot a}{2a}=0\,.
\ee
Here a dot denotes a derivative with respect to $u$.  It can be integrated and the resulting first-order equation is
\be
\dot a + \tilde\alpha\, a^2\log \big(\fft{a}{q}\big)=0\,,\qquad
\tilde \alpha = \ft12(n-1)\alpha\,.\label{firstorder}
\ee
where $q$ is an integration constant.  It is clear that $a=q$ is a stationary point of the equation, giving rise to the static solution presented earlier.  Interestingly, this black hole stationary point is stable.  For $a>q$, we have $\dot a<0$ whilst for $a<q$, we have $\dot a>0$.  The equation (\ref{firstorder}) can also be solved exactly in terms of an exponential integral function, namely
\be
{\rm Ei}\big(\log (\ft{q}{a})\big) = -\tilde\alpha q\,u\,.\label{aei}
\ee
We present the plot of the scalar charge $a$ as a function of the advanced time $u$
in Fig.~1.

\begin{figure}[ht]
\begin{center}
\includegraphics[width=200pt]{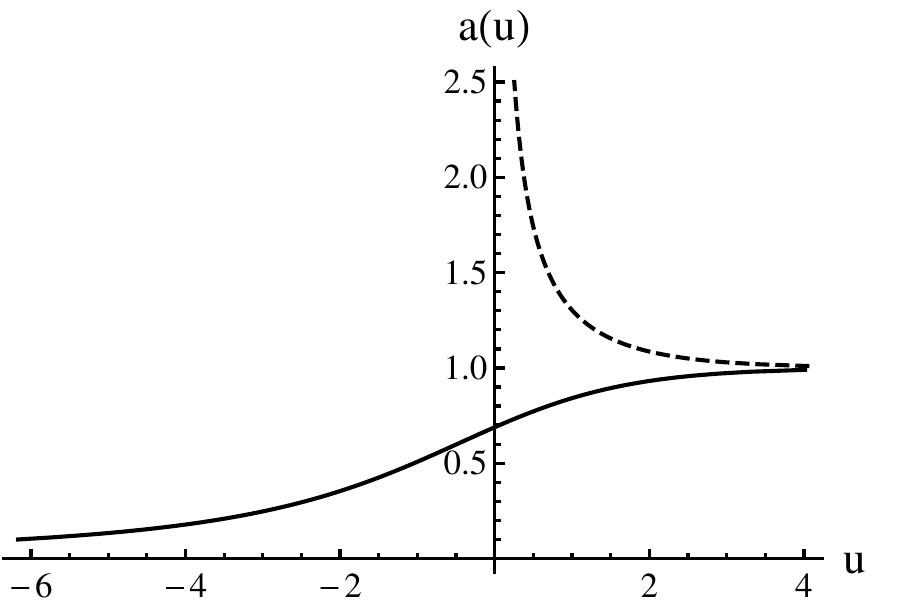}\ \
\end{center}
\caption{{\it The plot shows scalar charge $a(u)$ as a function of the advanced time coordinate $u$. The parameters $\tilde\alpha$ and $q$ are both set to unity.  There are two branches of the solution.  The solid line has $u\in (-\infty,\infty)$ whilst the dashed line has $u\in (0,\infty)$}}
\end{figure}

It is of interesting to note that there are two branches of the solutions, described by the solid and dashed lines in Fig.~1.  One way to understand the two branches is to note that as $u\rightarrow \infty$, we can perform the expansion
\be
a=q + c_0 e^{-\tilde \alpha q\,u} + \ft{3c_0^2}{2q} e^{-2\tilde \alpha q\,u} +
\ft{29c_0^3}{12q^2}  e^{-3\tilde \alpha q\,u} + \cdots\,,
\ee
where $c_0$ is an integration constant that can be absorbed by appropriate constant shift of $u$, up to the sign choice.  The positive and negative signs of $c_0$ correspond to the dashed and solid lines in Fig.~1 respectively.

The Vaidya mass \cite{Vaidya:1951zz} is given by (\ref{mass1}) with $q$ replaced by the dynamic scalar charge $a(u)$. Thus we have
\be
\dot M\equiv \fft{dM}{du}=\fft{(n-1)(n-2)^2\alpha^2}{256\pi} a^n \log\big(\fft{q}{a}\big)
\,.
\ee
It follows that $\dot M$ is positive for $a<q$, corresponding to the solid line in Fig.~1, and becomes negative for $a>q$, corresponding to the dashed line in Fig.~1.  For the dashed-line solution, the mass decreases with the advanced time $u$, and hence the solution should not be viewed as a white hole.  Instead, it absorbs ghost-like condensates.  To see this, we note that for any positive value of $a$, there exists an apparent horizon at $r=r_0(u)$ so that $f(r_0(u),u)=0$. For the solid-line solution, the apparent horizon is inside the event horizon and it approaches the event horizon as $u\rightarrow \infty$. On the other hand, for the dashed-line solution, the apparent horizon is outside the event horizon, indicating that the effective $\kappa(\phi)$ can be negative outside the event horizon, and hence the graviton modes become ghost-like in this region.  Usually, the existence of ghost modes suggests instability of a solution; however, it is rather interesting to note that the ghost modes here actually help to stablize the static black hole of $a=q$.

It should be emphasized that there are two stationary points for the first-order equation (\ref{firstorder}), namely $a=0$ and $a=q$.  The $a=0$ solution is the AdS vacuum (\ref{adsvac}) written in planar coordinates.  It is stable against perturbation at the linear level.  However, the nonlinear effect pushes the scalar charge $a=0$ from the past infinity time to $a=q$ at the future infinity.  At the past infinity $u\rightarrow -\infty$, the scalar charge behaves as $a\log a\sim -1/u$. During the whole dynamic process, regardless how small $a(u)$ is, an apparent horizon forms, which increases with advanced time $u$ and finally approaches the event horizon from within.  Although $u\in (-\infty, \infty)$, the dynamic process takes about semi-infinite times, since as $u$ runs from $-\infty$ to some positive region, the dynamic process starts to speed up and the solution approaches the static configuration exponentially with the relaxation time $\tau=1/(\tilde\alpha q)$.  Thus, the dynamic solutions we constructed provide analytical examples in general dimensions how the linearly-stable AdS vacua undergo nonlinear instability and evolve spontaneously into some stable black hole states.

Note that sometimes it is more convenient to use the scalar charge as the time coordinate.  To do this, we simply replace $du$ in the metric by
\be
du=\fft{1}{\tilde \alpha a^2 \log\fft{q}{a}}\,da\,.
\ee

We would like to point out that the hypergeometric functions in the solutions for the low-lying dimensions become much simpler and we present some explicit examples:
\bea
n=3:&&\cr
V&=&-\ft18g^2 - \ft14\alpha\Big(\fft{(2-\phi^2)\phi^2}{2(1-\phi^2)} +
\log (1-\phi^2)\Big)\,,\cr
f &=& g^2 r^2 - 2\alpha r \Big(a + r \log\big(1- \fft{a}{r}\big)\Big)\,;\cr
n=4:&&\cr
V&=&-\ft12g^2-\ft12\alpha\Big(\fft{\phi(3-2\phi^2)}{1-\phi^2} - 3{\rm arctanh}(\phi)\Big)\,,\cr
f&=&g^2 r^2 - 3 \alpha r \Big(-a + r {\rm arctanh}(\fft{a}{r})\Big)\,;\cr
n=5:&&\cr
V &=& -\ft98g^2-\ft38\alpha\Big(\fft{3\phi^{\fft23}(4-3\phi^2)}{1-\phi^2}+
2\log\fft{(1-\phi^{\fft23})^3}{1-\phi^2}- 4\sqrt3 \arctan\big(\fft{\sqrt3
\phi^{\fft23}}{2 + \phi^{\fft23}}\big)\Big)\,,\cr
f&=& g^2 r^2 - \ft23\alpha r\Big(-6a + 2\sqrt{3} r \arctan\big(\fft{\sqrt3\,a}{2r + a}\big) + r \log\fft{r^3-a^3}{(r-a)^3}\Big)\,;\cr
n=6:&&\cr
V &=& -2g^2 - \ft12\alpha\Big(\fft{4\sqrt{\phi}\,(5-4\phi^2)}{1-\phi^2}-10\arctan\sqrt{\phi}-10\,
{\rm arctanh}\sqrt{\phi}\Big)\,,\cr
f &=&g^2 r^2 - \ft54\alpha r\Big(-4a + 2r \arctan\big(\fft{a}{r}\big) + 2r\, {\rm arctanh}
\big(\fft{a}{r}\big)\Big)\,.
\eea
The scalar charge $a$ can either be constant $q$, giving rise to the static solution, or given by (\ref{aei}), describing dynamic black hole formation.

Finally we note that the above procedure of promoting a static solution to become dynamic one works for the static solution of case 2 only when $n=3$, which was already constructed in
\cite{Xu:2014xqa}.

\section{Einstein gravity with a minimally-coupled scalar}

We now turn our construction to Einstein gravity with a minimally-coupled scalar.  The Lagrangian is given  by
\be
{\cal L}_n= \sqrt{-g} \Big( R - \ft12 (\partial\phi)^2 - V(\phi)\Big)\,.
\ee
The equations of motion are
\be
E_{\mu\nu}\equiv G_{\mu\nu}- T_{\mu\nu}=0\,,\qquad
\Box\phi - \fft{dV}{d\phi}=0\,,
\ee
where
\be
T_{\mu\nu} = \ft12 \partial_\mu \phi \partial_\nu \phi - \ft12 g_{\mu\nu}
\big(\ft12(\partial\phi)^2 + V\big)\,.
\ee
Stationary black holes in these theories have been extensively constructed \cite{Anabalon:2012dw,Gonzalez:2013aca,Anabalon:2013sra,Acena:2013jya,Feng:2013tza}.  None of these solutions, except for the $D=4$ example \cite{Zhang:2014sta}, has been promoted to become dynamic solutions.  In this section, we construct more examples of static solutions and obtain a class of new solutions that can become dynamic.

\subsection{AdS planar black holes}

\subsubsection{Class 1}

In general $n\ge 3$ dimensions, we find that there exist a new class of exact planar black holes whose metric ansatz is
\be
ds^2 = - f dt^2 + \fft{dr^2}{f} + r(r + q) dx_i dx_i\,,\label{mini-ansatz-1}
\ee
where $f$ and also the scalar $\phi$ are functions of $r$ only, and $q$ is a constant. It turns out that the scalar $\phi$ can be solved directly, independent of the scalar potential.  We have
\be
e^{\lambda\phi} = 1 + \fft{q}{r}\,,\qquad \lambda=\sqrt{\ft2{n-2}}\,.
\ee
We can then use the equations of motion to determine both the function $f$ and the potential $V$, given by
\be f=r(r+q) \Big(g^2-\ft{\alpha q^{n-1}}{r^{n-1}}\,{}_2F_1(n-1,\ft n2,n,-\ft{q}{r})  \Big)\,,
\ee
and
\bea
V(\phi)&\!\!\!=\!\!\!&-\ft 12 g^2(n-2)\Big(n+(n-2)\cosh{(\lambda \phi)} \Big)-\ft 12 \alpha (n-2)(e^{\lambda\phi}-1)^{n-1}\cr
\!\!\!\!&\times&\!\!\!\Big((n-1)(1+e^{-\lambda\phi})e^{-\fft{\phi}{\lambda}}- \Big[n+(n-2)\cosh{(\lambda \phi)} \Big]{}_2F_1(n-1,\ft12 n,n,1-e^{\lambda\phi}) \Big).
\eea

   To analyse the asymptotic structure of the solution, we define the luminosity distance $R=\sqrt{r(r+q)}$.  At the asymptotic large $R$, we have
\be
f=g^2R^2 - \fft{\alpha q^{n-1}}{R^{n-3}} + \fft{(n^2-1)\alpha q^{n+1}}{8 R^{n-1}} + \cdots\,.
\ee
It follows that the mass of the solution is given by
\be
M=\fft{(n-2)\alpha q^{n-1}}{16\pi}\,.\label{mass2}
\ee
Since $f(0)\rightarrow -\infty$, there must be an horizon at $r=r_0$ where $f(r_0)=0$.  The temperature and entropy density are given by
\be
T=\ft{1}{4\pi} (n-1)\alpha q^{n-1}\big(r_0(r_0+q)\big)^{1-\fft12n}\,,\qquad
S=\ft14 \big(r_0(r_0+q)\big)^{\fft12n-1}\,.
\ee
It is straightforward to verify that both the first law (\ref{firstlaw}) and the Smarr relation (\ref{smarr}) are satisfied.

   Since $n$ is an integer, the hypergeometric function becomes much simpler.  For example, when $n=3$, the solution and the corresponding scalar potential are given by
\bea
&&f=(g^2+8\alpha)r(r+q)-4\alpha(2r+q)\sqrt{r(r+q)}\,,\cr
&&V(\phi)=-\ft 12g^2\Big(3+\cosh{(\sqrt{2}\phi)}\Big)-32\alpha\sinh^4{(\ft{\phi}{2\sqrt{2}})}\,.
\eea
For $n=4$ dimensions, we find
\bea
&&f=g^2r(r+q)-3\alpha q^2-6\alpha q r+6\alpha r(r+q)\log{(1+\frac{q}{r})}\,,\cr
&&V(\phi)=-2g^2(2+\cosh{\phi})-12\alpha \Big(\phi(2+\cosh{\phi})-3\sinh{\phi}  \Big)
\,.
\eea
This solution with negative $g^2$ was constructed in the context of cosmology in
\cite{Zloshchastiev:2004ny}. For $n=5$ dimensions, we find
\bea
&&f=(g^2-\ft{128}{3}\alpha) r(r+q)+\frac{8\alpha(2r+q)(8r^2+8qr-q^2)}{3\sqrt{r(r+q)}}\,,\cr
&&V(\phi)=-\ft32 g^2\Big(5+3\cosh{(\sqrt{\ft 23}\phi)}\Big)-1024\alpha\sinh^6{(\ft{\phi}{2\sqrt{6}})}\,.
\eea

The anasatz (\ref{mini-ansatz-1}) is a special case of a more general class of ans\"atze that one can consider, namely
\be
ds^2=-f dt^2 + \fft{dr^2}{f} + r^{1+\mu} (r+q)^{1-\mu} dx^i dx^i\,.
\ee
It can be cast into the form
\be
ds^2=\Omega(x) \Big(-h(x) dt^2 + \fft{dx^2}{h(x)} + dx^i dx^i\Big)\,,
\ee
with
\be
\Omega(x)=\fft{q^2 x^{\nu-1}}{(1-x^{\nu})^2}\,,\qquad \nu=\fft{1}{\mu}\,.
\ee
This ansatz was first considered in \cite{Anabalon:2012ta}.  The higher dimensional planar black holes for a generic finite $\nu$ was obtained \cite{Acena:2013jya}.
Our $\mu=0$ solution is thus a non-trivial singular limit $\nu\rightarrow \infty$ of the general class of solutions constructed in \cite{Acena:2013jya}.

\subsubsection{Class 2}

We also find that the theory admits a different class of new planar black holes in $n$ dimensions with the metric ansatz given by:
\be
ds^2=-\sigma^2 fdt^2+\frac{dr^2}{f}+r^2dx_i^2\,,
\ee
where the functions $f$, $\sigma$ and the scalar $\phi$ all depend on $r$ only.

From the Einstein equation (E$^t_{\,t}$-E$^r_{\,r}$), we find that
\be
\phi'^2=\frac{2(n-2)\sigma'}{r \sigma}\,.\label{phieom}
\ee
We would like that the scalar vanishes at the asymptotic infinity.  The following might be the simplest ansatz satisfying this criterium:
\be
\phi= \big(\fft{q}{r}\big)^\mu\,,
\ee
where $\mu>0$ and the constant $q>0$ may be viewed as the scalar charge.  The equation (\ref{phieom}) implies that
\be
\sigma=e^{-\lambda^2 \phi^2}\,,\qquad
\lambda=\sqrt{\ft{\mu}{4(n-2)}}\,.\label{sigsol}
\ee
The function $f$ can then be determined from one of the combinations of the Einstein equations and we find
\be
f=e^{2\lambda^2\phi^2} r^2 \Big(g^2 + \alpha\,\nu\,\lambda^{-2\nu} \big(\Gamma(\nu, \lambda^2 \phi^2)-\Gamma(\nu)\big)\Big)\,,\qquad \nu=\ft{n-1}{2\mu}\,.\label{fsol}
\ee
The scalar potential responsible for this solution is given by
\bea
V &=&-2g^2(n-2)\mu (\nu-\lambda^2\phi^2)e^{2\lambda^2\phi^2}\cr
&& -(n-1)(n-2)\alpha e^{\lambda^2\phi^2} \Big(\phi^{2\nu} + \lambda^{-2\nu} (\nu-\lambda^2\phi^2)
e^{\lambda^2\phi^2}\big(\Gamma(\nu,\lambda^2\phi^2)-\Gamma(\nu)\big)\Big)\,.
\eea
At the asymptotic region $r\rightarrow \infty$ region, we have
\be
-g_{tt}=\sigma^2 f = g^2 r^2 -  \fft{\alpha q^{n-1}}{r^{n-3}} +
\fft{\alpha \mu(n-1)q^2}{4(n-2)(n+2\mu-1)}\big(\fft{q}{r}\big)^{n+2\mu-3}+ \cdots\,,
\ee
It follows that the mass has the same form as (\ref{mass2}).  It can also be easily established that there is a horizon at $r=r_0>0$.  The temperature and entropy are given by
\be
T=\ft{1}{4\pi} (n-1)\alpha q\, \big(\fft{q}{r_0}\big)^{n-2}\,,\qquad
S=\ft14 r_0^{n-2}\,.
\ee
It is easy to verify that the first law of thermodynamics and the Smarr relation are both satisfied.

\subsection{Dynamic solutions}

Having obtained the two classes of static AdS planar black holes involving a scalar charge, we shall now try to promote the scalar charge to be time-dependent. As in the previous examples involving a non-minimally-coupled scalar, we first write the static solutions in the Eddington-Finkelstein-like coordinates. For the first class, we have
\be
ds^2 = 2 du dr - f du^2 + r(r+q) dx^i dx^i\,.
\ee
It turns out that sending $q\rightarrow a(u)$ will not solve the full set of equations of motion except in $n=4$ dimensions.  The four-dimensional dynamic solution was obtained in \cite{Zhang:2014sta}, describing the exact formation of some scalar hairy black hole.

Analogously, for the second class of solutions, the time-dependent ansatz is
\be
ds^2 = -2\sigma du dr - \sigma^2 f du^2 + r^2 dx^i dx^i\,,\qquad
\phi=\big(\fft{a(u)}{r}\big)^{\mu}\,,
\ee
where the functions $\sigma$ and $f$ take the same forms as (\ref{sigsol}) and (\ref{fsol}) respectively.  Note that we have adopted retarded time $u$ in the metric for this case. Substituting the ansatz into the equations of motion, we find that a condition for having a time-dependent solution is
\be
\mu=\ft12(n-2)\,.
\ee
Consequently, the full set of equations of motion can be satisfied provided that
\be
\dot a (\dot a +\tilde \alpha\, a^2)=0\,,\qquad \tilde \alpha =
\ft{4(n-1)\alpha}{n-2}\,.
\ee
Thus we have
\be
a=q\,,\qquad \hbox{or}\qquad a=\fft{1}{\tilde \alpha\, u}\,.
\ee
The static solution is marginally stable against linear perturbation.  The dynamic solution describes a radiating white hole that eventually become the AdS vacuum in planar coordinates at late retarded times.  The result indicates that the AdS vacuum may be stable at the full nonlinear level.

\section{Conclusions}

In this paper, we considered Einstein gravities in general dimensions, coupled to a scalar field either minimally or non-minimally, together with a generic scalar potential.  By choosing the scalar potential appropriately, we obtained large classes of black hole solutions that are asymptotic to AdS spacetimes in planar coordinates.  These planar black holes contain a scalar charge $q$ as a free integration constant.  The mass, temperature and entropy are all functions of $q$, satisfying the first law of thermodynamics $dM=TdS$.  Since the Schwarzschild AdS planar black holes are also solutions of these theories, the black hole uniqueness is no longer valid and the new black holes are scalar hairy.  Note also that the (massive) scalar in our theories satisfies the Breitenlohner-Freedman bound and hence the AdS vacua are all linearly stable.

We then followed \cite{Zhang:2014sta} and wrote all the static solutions in Eddington-Finkelstein coordinates and promoted the non-conserved scalar charge $q$ to be dependent on the advanced or retarded time coordinate $u$.  For the majority of such ans\"atze, the equations of motion cannot be satisfied.  However, we obtained two classes of exact dynamic solutions: one describes black hole formation whilst the other describes radiating white holes.  The black hole formation solution is of particular interest, with the advanced time coordinate $u$ running from $-\infty$ to $+\infty$.  During this period, the spacetime evolves from the AdS vacuum to a stable black hole state, with the apparent horizon grows from zero at the infinity past to coincide with the event horizon eventually.  Since the scalar driving the dynamic process is conformally massless, the AdS vacuum is linearly stable.  It is thus due to the nonlinear effects that the planar black hole forms spontaneously from the AdS vacuum.

    Our solutions of dynamic black hole formation thus provide an explicit and analytical demonstration of the nonlinear instability of the linearly-stable AdS vacua.

\section*{Acknowledgement}

 Z.-Y.~Fan is supported in part by NSFC Grants NO.10975016, NO.11235003 and NCET-12-0054; The work of H.L.~is supported in part by NSFC grants 11175269, 11475024 and 11235003.

\end{document}